\documentclass[preprintnumbers,aps,pre,twocolumn,superscriptaddress,amsmath,amssymb]{revtex4}

\usepackage{color}
\usepackage{graphicx}
\usepackage{dcolumn}
\usepackage{bm}
\usepackage{floatrow}

\newcommand{\Qvec}{\mathbf{Q}}
\newcommand{\nvec}{\mathbf{n}}
\newcommand{\Tr}[1]{\mathrm{Tr} #1}

\begin{document}

\author{Z. Eskandari}
\affiliation{Department of Physics, Sharif University of Technology, P.O. Box: 11155-9161, Tehran, Iran.}

\author{N. M. Silvestre}
\email{nunos@cii.fc.ul.pt}
\affiliation{Centro de F{\'\i}sica Te\'orica e Computacional,
Universidade de Lisboa, Avenida Professor Gama Pinto 2, PT-1649-003 Lisboa, Portugal}
\affiliation{Departamento de F{\'\i}sica, Faculdade de Ci\^encias, Universidade de Lisboa, Campo Grande, PT-1749-016 Lisboa, Portugal.}

\author{M. M. Telo da Gama}
\affiliation{Centro de F{\'\i}sica Te\'orica e Computacional,
Universidade de Lisboa, Avenida Professor Gama Pinto 2, PT-1649-003 Lisboa, Portugal}
\affiliation{Departamento de F{\'\i}sica, Faculdade de Ci\^encias, Universidade de Lisboa, Campo Grande, PT-1749-016 Lisboa, Portugal.}

\author{M. R. Ejtehadi}
\affiliation{Department of Physics, Sharif University of Technology, P.O. Box: 11155-9161, Tehran, Iran.}

\title{Particle selection through topographic surface patterns in nematic colloids}

\begin{abstract}
We propose the use of topographic modulation of surfaces to select and localize particles in nematic colloids. By considering convex and concave deformations of one of the confining surfaces we show that the colloid-flat surface repulsion may be enhanced or switched into an attraction. In particular, we find that when the colloidal particles have the same anchoring conditions as the patterned surfaces, they are strongly attracted to concave dimples, while if they exhibit different anchoring conditions they are pinned at the top of convex protrusions. Although dominated by elastic interactions the first mechanism is reminiscent of the depletion induced attraction or of the key-lock mechanism, while the second is specific to liquid crystal colloids. These long-ranged, highly tunable, surface-colloid interactions contribute for the development of  template-assisted assembly of large colloidal crystals, with well defined symmetries, required for applications.

\end{abstract}

\maketitle

\section{Introduction}

Colloidal systems are an important class of metamaterials \cite{Xia.2000,Liu.2011}, having unusual optical and mechanical properties. Assembling them into mesoscale structures with controlled symmetries is a primary challenge for their use in a wide variety of applications, including responsive devices \cite{Ge.2011} and photonic band-gap crystals \cite{Zografopoulos.2012}. Among the techniques to grow large scale three-dimensional colloidal crystals \cite{Xia.2000,Vickreva.2000,Crane.2013} the use of topographic templates, to assist the assembly of colloidal particles, has been shown to effectively control the symmetry and orientation of colloidal crystals \cite{Blaaderen.1997,Yin.2001,Wang.2004,Rycenga.2009,Sharma.2011}. This method relies on topographically, or chemically, patterned surfaces, tuned through conventional micro-fabrication techniques such as soft lithography, as templates to drive the nucleation and epitaxial growth of colloidal structures \cite{Yin.2002}. 
Template assisted assembly offers advantages over conventional methods, namely (i) templated structures have fewer defects and longer ranged order and (ii) novel structures may be assembled with the tunable symmetry of the template pattern \cite{Varghese.2006}. Templates with patterns commensurate with those of the colloids are also useful in applications involving size selectivity \cite{Varghese.2006}. In the context of recent advances on the fabrication of anisotropic colloidal particles, both in shape and chemical surface patterns \cite{Lee.2011}, particle selectivity becomes increasingly important, in terms of the colloidal size and shape as well as in terms of the colloidal surface chemistry.

Colloidal particles in simple fluids exhibit short range interactions, e.g., depletion forces in isotropic solvents. To facilitate the assembly process suitable techniques have been employed, depending on the type of interaction between the colloids and the template surface \cite{Xia.2000,Vickreva.2000,Varghese.2006}. 
In liquid crystal (LC) colloids, particles with sizes ranging from a few hundred nanometers to a couple of ten micrometers interact via the elastic distortions of the LC matrix.
The surface of each colloid induces distortions in the LC orientational field, leading to the nucleation of topological defects, which result in long-range anisotropic interactions between colloidal particles, with binding potentials of the order of several hundred to a few thousand $k_BT$ \cite{Loudet.2000,Musevic.2006,Ognysta.2008,Tasinkevych.2012,Nych.2013}. 
 As a consequence, the colloids self-assemble into colloidal structures that are highly dependent on the symmetry of their pair-wise interaction \cite{yada.2004,Ravnik.2007}. 
Typically, colloidal particles in nematic LCs exhibit short-distance repulsion that prevents them from touching, and the equilibrium distance between colloids can be tuned through the application of external fields \cite{Loudet.2001,Noeel.2006}. 
 Combining the mechanical response with the electro-optical properties of the LC host makes LC colloids suitable candidates for responsive photonic devices \cite{Musevic.2011}.
 
In LC colloids the surface chemical treatment determines how the LC molecules will align (anchor) near the colloidal surface. Typically, LC colloids exhibit either homeotropic (perpendicular) or planar (parallel) surface anchoring \cite{Poulin.1998}. By adding photosensitive azobenzene groups to the surface of the colloidal particles it is possible to switch between anchoring conditions \cite{Chandran.2011}. Janus colloids with different types of anchoring on each side have also been produced  \cite{Conradi.2009}.  

Typically, colloidal particles exhibit a strong repulsion from the (flat) cell walls that pushes them towards the middle of the cell. Previous studies of nematic colloids with homeotropic anchoring near surfaces with the same anchoring conditions, i.e., homeotropic, revealed that such repulsion is turned into a strong attraction if the confining surface exhibits concave topographic patterns with a shape complementary to that of the colloidal particle \cite{Silvestre.2004,Hung.2007}. The trapping potential of this key-lock mechanism is $\sim400k_BT$ for sub-micrometer colloids \cite{Hung.2007}. Recently, it was pointed out that when the flat surface is decorated with spherical (convex) pillars the repulsion may also be changed into an attraction if the surface anchoring conditions are different, providing a mechanism for the assembly of colloidal particles by surface topographic patterns \cite{Eskandari.2013}.

In the following, we explore the use of topographic patterned surfaces for particle selectivity based on their surface chemistry (anchoring). By combining the surface topography (concave or convex) and the anchoring conditions, we show how 
the repulsion from flat surfaces may be turned into an attraction or further enhanced, providing a mechanism for particle selection in nematic colloids based on their surface chemistry.

This paper is organized as follows. In the next section we describe the characteristics of the topographic patterns considered and the model used to describe the nematic LC host. In section III we describe and analyze the results. In particular, we show that when the colloidal particles and the template have the same anchoring conditions, concave patterns induce an attraction while convex patterns enhance the repulsion between the colloids and the confining surface. In addition, when the colloidal particles and the template have distinct anchoring conditions, the phenomenon is reversed and the colloids are attracted towards the convex patterns and repelled from the concave ones. Finally, in section IV we present our conclusions.

\section{Model}

\begin{figure}[t]
\centerline{\includegraphics[width=.6\columnwidth]{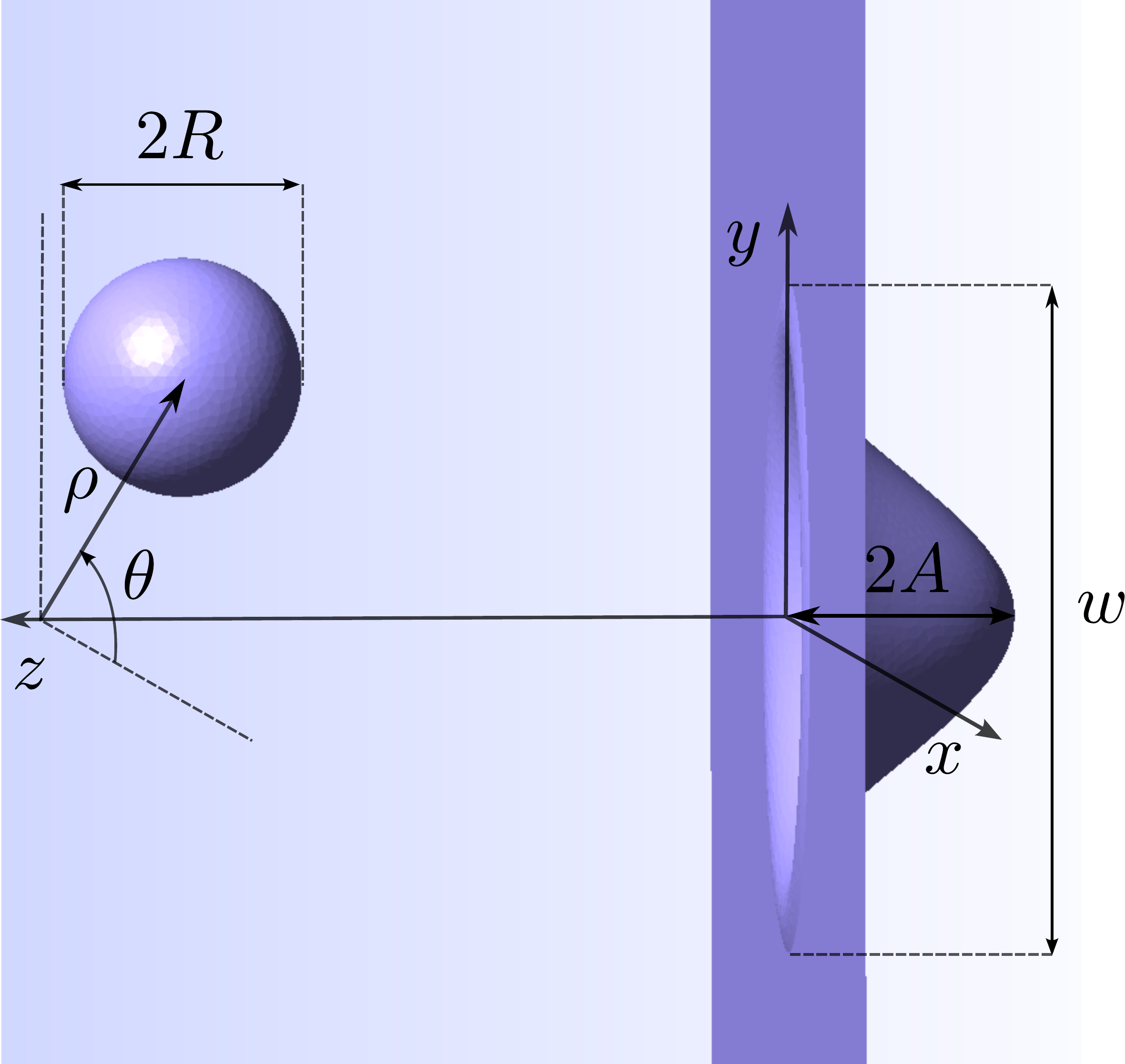}}
\caption
{(color online) Schematic representation of a colloid of radius $R$ close to a topographically patterned surface. The pattern is a sinusoidal deformation, with cylindrical symmetry, modelled as $A(1+\cos{(2\pi \rho/w)})$. The width  and depth of the pattern are $w$ and $2A$, respectively. $A>0$ corresponds to a convex protrusion on the surface into the nematic cell, while $A<0$ models a concave dimple. The polar coordinates on a plane perpendicular to the surface are denoted $(\rho,\theta)$.}
\label{geo}
\end{figure}

In this paper we consider the interaction of nematic colloids with convex (protrusions) and concave (dimples) topographic patterns. The patterns are modelled by local sinusoidal deformations of the surface, as sketched in Fig.\ref{geo}, with cylindrical symmetry, described by $A(1+\cos{(2\pi \rho/w)})$. $\rho$ is the distance to the center of the pattern. For convenience the size of the deformations are comparable to the radii of the colloids under consideration, $R$. The width and depth of the patterns are $w=2.2R$ and $2A$, respectively. Convex protrusions are described by $A>0$, indicating that the pattern points into the nematic cell, while concave dimples are described by $A<0$.

The nematic LC is modelled by the Landau-de Gennes (LdG) mean-field theory \citep{Gennes.1993}, where the local orientational order is described by a tensor order parameter $\Qvec$, traceless and symmetric, which for a uniaxial nematic is $Q_{ij}=S(3n_in_j-\delta_{ij})/2$. $\nvec$ is the nematic director field and $S$ is the scalar order parameter (degree of order). The free energy is written in terms of $\Qvec$ and its derivatives,
 $F=\int_V{d^3x\left(f_b\left(Q_{ij}\right)+f_e\left(\partial_kQ_{ij}\right)\right)}+\int_{\partial V}{ds f_s\left(Q_{ij}\right)}$, where the bulk and elastic free energy densities are, respectively,
 \begin{eqnarray}
 f_b&=&a_o\left(T-T^*\right)\Tr \Qvec^2-b\Tr\Qvec^3+c\left(\Tr\Qvec^2\right)^2\\
 f_e&=&\frac{L_1}{2}\partial_kQ_{ij}\partial_kQ_{ij}.
 \end{eqnarray}
For simplicity, we consider the one-elastic constant approximation. For 5CB in the nematic phase the LdG parameters are \cite{Kralj.1991} $a_o=0.044$ MJ K$^{-1}$ m$^{-3}$ , $b=0.816$ MJ m$^{-3}$, $c=0.45$ MJ m$^{-3}$, $L_1=6$ pJ m$^{-1}$, and $T^*=307$ K. 
We have set the temperature $T=307.2$ K, just below the isotropic-nematic transition temperature, $T_{\mathrm{IN}}(\mathrm{5CB})=308.5$K.

The surface contribution to the free energy arises from the anchoring on the LC molecules. The anchoring conditions on the colloids and the template may be either homeotropic or planar (degenerate).
For homeotropic anchoring, we use the Nobili-Durand surface potential \cite{Nobili.1992} $f_s=W\Tr(\Qvec-\Qvec^s)^2$ that penalizes deviations from the preferred surface order parameter $Q^s_{ij}=S_b(3\nu_i\nu_j-\delta_{ij})/2$. $S_b=(b/8c)(1 + \sqrt{1 − 8\tau/9})$ is the equilibrium scalar order parameter in the bulk, where $\tau=24 a c/b^2$, and {\boldmath$\nu$} is the normal to the surface. 
For planar degenerate anchoring, we use the Fournier-Galatola surface potential \cite{Fournier.2005} $f_s=W\left(\Tr\left(\tilde{\Qvec}-\tilde{\Qvec}^\perp\right)^2
+\left(\Tr\tilde{\Qvec}^2-3S_b^2/2\right)^2\right)$, where $\tilde{Q}_{ij}=Q_{ij}-S_b/2$, and  $\tilde{Q}_{ij}^\perp=\left(\delta_{ij}-\nu_i \nu_j\right) \tilde{Q}_{ij}\left(\delta_{ij}-\nu_i \nu_j\right)$. A discussion of the effect of the quadratic and the quartic terms may be found in \cite{Tasinkevych.2012}.

We minimize the Landau-de Gennes free energy numerically, using finite elements (FEM) with adaptive meshing coupled with the conjugate gradient (CG) method, with a precision better than $1\%$. Details of the numerics may be found in \citep{Tasinkevych.2012}.

\section{Results and discussion}

\subsection{Homeotropic colloid}
\subsubsection{Homeotropic surface}

In a flat cell, a sub-micrometer spherical colloid with homeotropic anchoring, in an otherwise uniform nematic, with a Saturn-ring line defect finds itself in equilibrium at the center of the cell due to repulsive forces from the surfaces, which decay as $1/z^6$ \cite{Chernyshuk.2011}, where $z$ is the distance from the surface. This repulsion, arising from the elastic distortion of the nematic matrix, is modified by topographic modulations of the confining surfaces. 

\begin{figure}[t]
\centerline{\includegraphics[width=0.9\columnwidth]{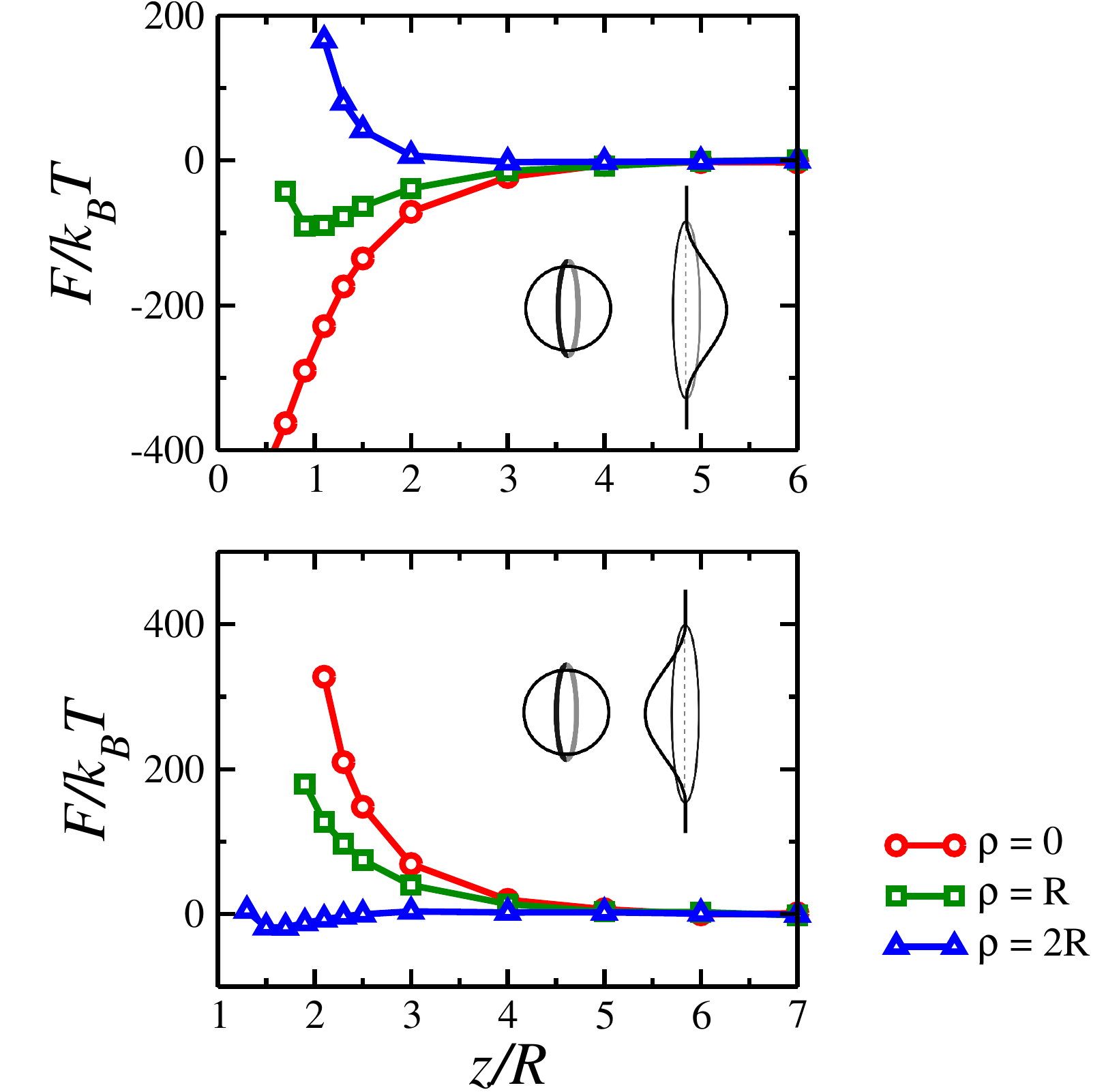}}
\caption{(color online) Interaction free energy of a colloid 
of radius $R$ with (up) a concave and (down) a convex topographic pattern, with homeotropic surface anchoring everywhere, as a function of the distance $z$ to the flat surface. 
$\rho$ is the distance between the cylindrical symmetry axis and the center of the colloid, defined in Fig.\ref{geo}.}
\label{perp}
\end{figure}

When the surface itself induces homeotropic anchoring, a concave topographic deformation will change the flat-surface repulsion, into an attraction. This phenomenon, reported previously in two \cite{Silvestre.2004} and  three dimensions \cite{Hung.2007}, is driven by shape complementarity known as the key-lock mechanism. Figure \ref{perp} depicts the interaction of a colloidal particle with a concave (up) and a convex (down) surface as function of the distance $z$ to the reference plane, and several distances from the symmetry axis $\rho=0,\,R,\,$ and $2R$. For a concave topographic pattern (dimple), the colloidal particle exhibits a strong attraction and the dimple will capture the colloid, with a binding potential $\sim 500k_BT$. As the colloid is moved away from the dimple axis of symmetry, the elastic stress pulls the colloidal particle towards its center while, at the same time, pushes the colloid away from the flat surface. This is consistent with previous results for cavities with cylindrical symmetry as well as for channels \cite{Silvestre.2004,Hung.2007}. The effect of a protrusion, however, is quite different from that of a dimple. 

In the presence of a protrusion with homeotropic anchoring the colloidal particle is pushed away from the topographic pattern. There is a very weak attraction towards the base of the protrusion ($\rho=2R$), where it meets the flat surface, with a binding potential $~18k_BT$. This is reminiscent of the interaction between colloidal particles with homeotropic anchoring, which attract at oblique angles with respect to the global director orientation. Here the protrusion plays the role of the second colloid. However the repulsion from the flat surface weakens the attraction at oblique angles.
The main feature of the convex topography is the enhancement of the repulsion from flat surfaces. This is clearly seen in Fig.\ref{A}$a)$ that illustrates how the interaction of homeotropic colloids with the topographic pattern changes with its height/depth $2A$. Dimples ($A<0$) with depth comparable to the size of the colloid exhibit strong attractive potentials. As the depth decreases and the dimple turns into a convex protrusion, the strength of the attractive potential decreases and changes into a repulsion that increases with the protrusion height.
These results indicate that in the presence of these types of topographic patterns with homeotropic anchoring, colloids with Saturn-ring defects will assemble at the dimples rather than at the protrusions.

\subsubsection{Planar wall}

\begin{figure*}[t]
\centerline{\includegraphics[width=0.9\columnwidth]{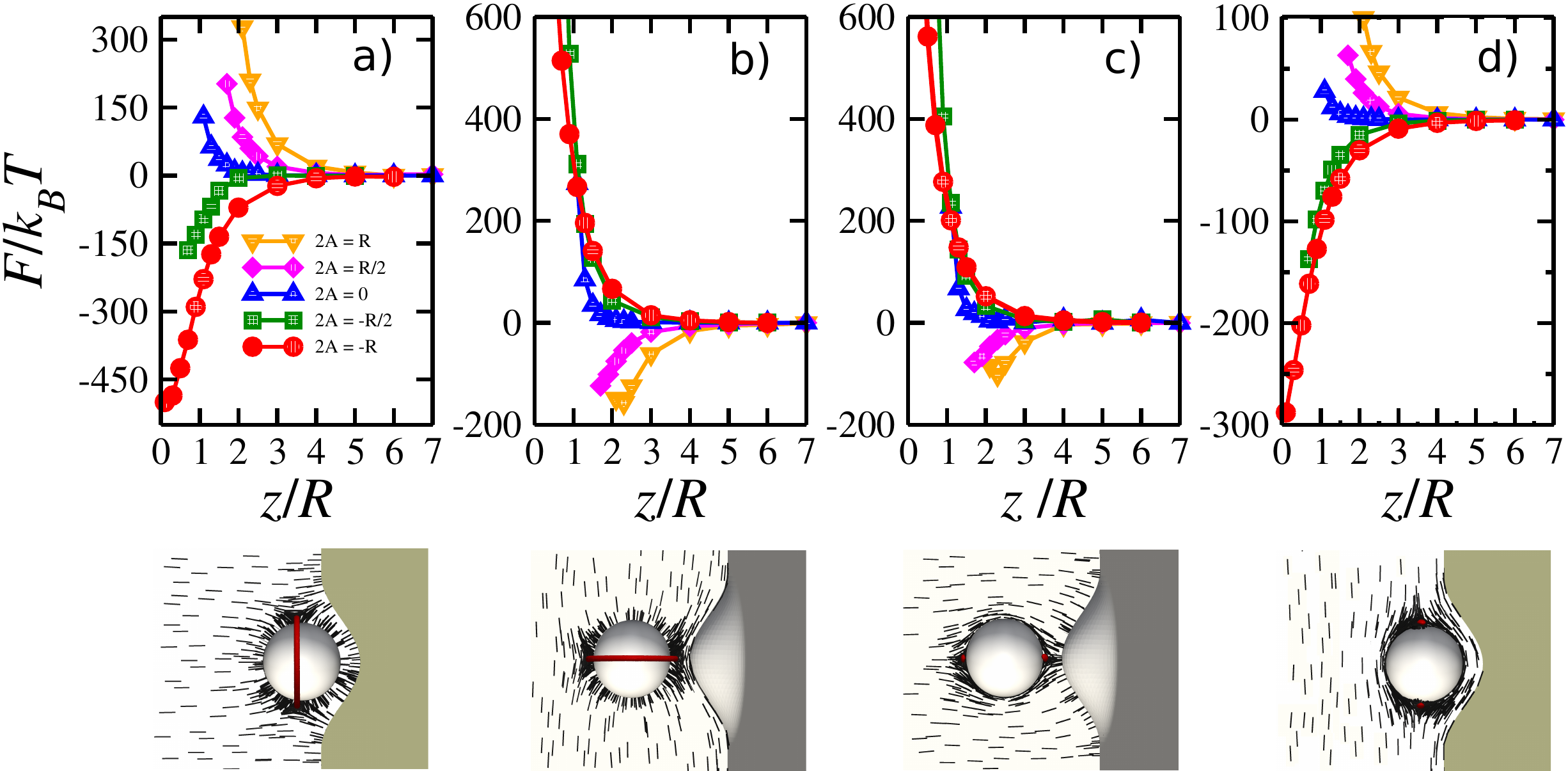}}
\caption{(color online) Interaction free energy of a colloid with homeotropic, $(a)$ and $(b)$, and planar anchorings, $(c)$ and $(d)$, with topographic patterns with homeotropic, $(a)$ and $(c)$, and planar anchorings, $(b)$ and $(d)$, as function of the distance to the surface. We consider convex protrusions with heights $2A=R/2,\,R$ and concave dimples with depths $2A=-R/2,\,-R$. For comparison the interaction with the flat surface $2A=0$, at $d=0$, is also represented. The insets illustrate the anchoring conditions at each surface, and depict equilibrium configurations.}
\label{A}
\end{figure*}

When the anchoring at the topographically patterned surface is planar the situation is reversed.
Figure \ref{A}$b)$ depicts the interaction of a colloid with homeotropic anchoring with a topographic pattern with planar degenerate anchoring, as a function of the distance to the wall $z$. The global orientation of the nematic director is parallel to the surface. As a result, the defect nucleated around the colloid is perpendicular to the surface. By contrast to homeotropic surfaces, there is a strong repulsion from concave dimples and an attraction towards convex protrusions, particularly towards the tips, with binding potentials of a couple of hundred $k_BT$. As a result, for antagonistic anchoring conditions, homeotropic colloids assemble at the top of convex protrusions. This is again reminiscent of the interaction between two colloidal particles with mismatching anchoring conditions. Both experiments \cite{Ognysta.2011} and numerical calculations \cite{Eskandari.2012} indicate that such colloids assemble either along the nematic director or perpendicular to it. Here the convex protrusion plays the role of a colloid with planar anchoring, and the homeotropic colloid is attracted to the tip of the convex topographic pattern, perpendicular to the global director orientation.

\subsection{Planar colloid}
The question arises whether similar phenomena may be observed for colloidal particles with planar anchoring. Such colloids exhibit two surface defects, known as {\it boojums}, located at antipodal positions and aligned with the global director orientation, and are known as {\it boojum-colloids} \cite{Tasinkevych.2012}. It turns out that boojum-colloids, in the presence of topographically patterned surfaces with homeotropic anchoring (see Fig.\ref{A}$c)$), assemble at the top of convex protrusions, with a binding potential of the order of $100k_BT$. This is similar to what was found for homeotropic colloids near planar surfaces. The homeotropic concave dimple enhances the repulsion between the boojum-colloid and the surface, while the homeotropic convex protrusion plays the role of a homeotropic colloid; the boojum-colloid is attracted to  its tip along the global director orientation.

\begin{figure*}[t]
\caption{Interaction energy of a boojum-colloid with a dimple (up) and a protrusion (down) with planar anchoring, as a function of the distance $z$ to the surface, for different polar orientations $\theta=0,\,45^\circ,\,90^\circ$, defined in Fig.\ref{geo}, and distances from the (cylindrical) symmetry axis $\rho=0$ (circles), $R$ (squares), and $2R$ (triangles). The depth/height of the pattern is $|2A|=R$.}
\label{planar_energy}
\centerline{\includegraphics[width=1.0\columnwidth]{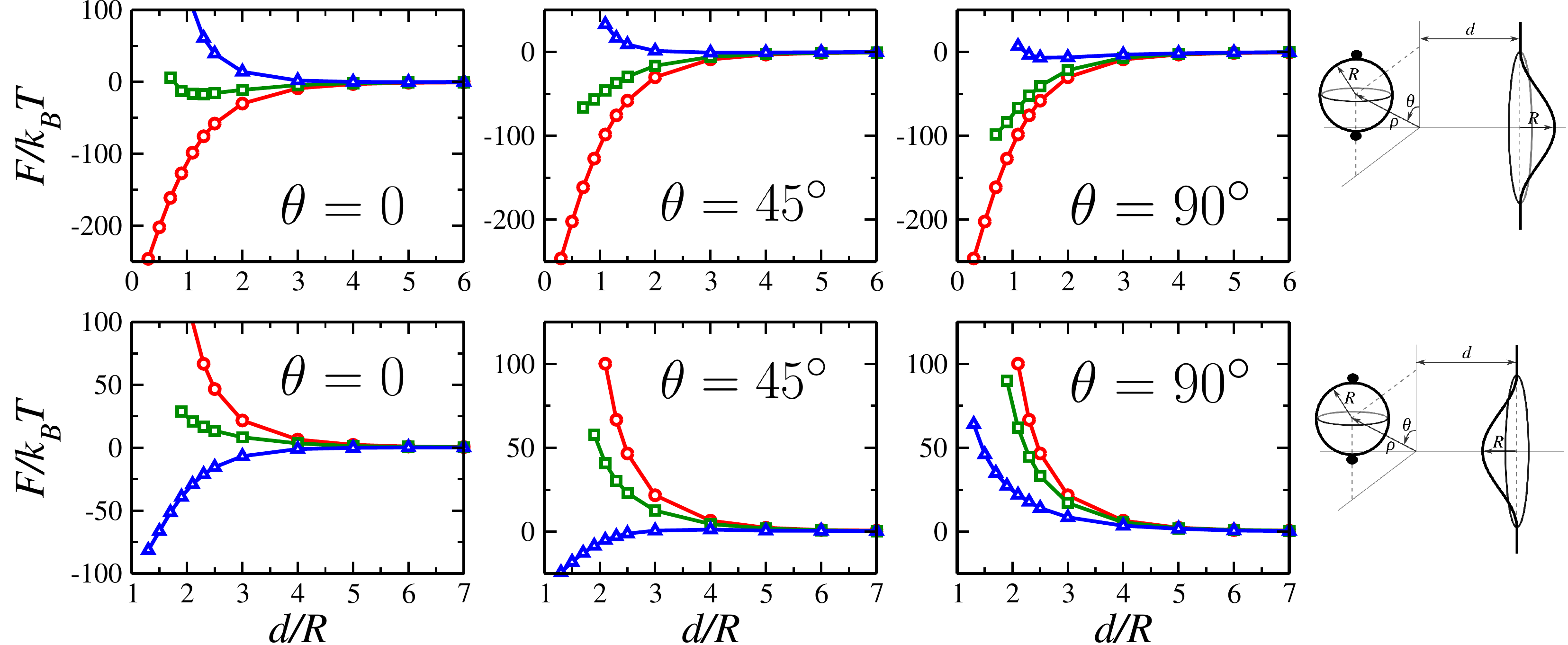}}
\end{figure*}

At surfaces with planar anchoring (Fig.\ref{A}$d)$), the situation is changed and the boojum-colloid assembles at concave dimples with a binding potential of $\sim 300k_BT$. In this case, the concave topography plays the role of a lock and the boojum-colloid that of a matching key. Similar to previous results for the key-lock mechanism with homeotropic anchoring \cite{Silvestre.2004,Hung.2007}, Fig\ref{planar_energy} (up) illustrates that there is a cone of attraction that will drive the colloidal particle to the concave dimple, and outside of which ($\rho\geq 2R$) the interaction between the colloid and the surface is purely repulsive. Figure \ref{planar_energy} (down) also shows that for planar convex protrusions boojum-colloids may be trapped at an oblique angle to the symmetry axis of the pattern. This is again related to the fact that convex protrusions play the role of a secondary colloid.

\section{Conclusions}

We investigated the interaction of nematic colloids with topographically patterned surfaces. We considered both the colloid and the surface with homeotropic or planar (degenerate) anchoring and topographic patterns with sizes comparable to the colloids.
For concave topographies and matching anchoring conditions the interaction is reminiscent of the key-lock mechanism,   
while for convex ones and distinct anchorings the interaction is similar to that between quadrupolar colloids, with opposite quadrupolar moments. 

Our findings reveal a robust selectivity mechanism through the localization of distinct colloids by topographic surface patterns. Colloidal particles with homeotropic anchoring assemble at concave dimples if, and only if, the surface has homeotropic anchoring, and at the top of convex protrusions if the surface has planar anchoring. The situation is reversed for colloidal particles with planar anchoring. In this case the colloids assemble at concave dimples if the anchoring at 
the surface matches that of the colloid, and at the top of the protrusions if there is a mismatch.

This work reveals that topographic modulation of surfaces provides the basis for selectivity processes based on the matching, or mismatching, of the colloids and surface anchoring. Convex or concave topographic patterns, or a combination of both, with a prescribed distribution will drive the controlled assembly of colloidal particles, or even contribute to the fabrication of new optical devices.

The results of this work may be extended to concave channels and convex ridges that could be used as hard rails to select and, in the presence of weak flows, divert colloidal particles to specific regions in microfluidic chips.

\section*{Acknowledgements}

We acknowledge the financial support from the Portuguese Foundation for Science and Technology (FCT) through grants n.o PEstOE/FIS/U10618/2014, PTDC/FIS/098254/2008, EXCL/FIS-NAN/0083/2012, and SFRH/BPD/40327/2007 (NMS).


\end{document}